\documentclass[prl,twocolumn,superscriptaddress,showpacs,preprintnumbers,amsmath,amssymb,floats]{revtex4}

\usepackage{txfonts}
\usepackage{amssymb}
\usepackage{graphicx}

\begin{document}

\title{Orbital selective correlations between nesting/scattering/Lifshitz transition and the superconductivity in  AFe$_{1-x}$Co$_x$As (A=Li, Na)}

\author{Z. R. Ye}
\author{Y. Zhang}
\author{M. Xu}
\author{Q. Q. Ge}
\author{Q. Fan}
\author{F. Chen}
\author{J. Jiang}
\affiliation{State Key Laboratory of Surface Physics, Department of
Physics,  and Advanced Materials Laboratory, Fudan University,
Shanghai 200433, People's Republic of China}

\author{P. S. Wang}
\author{J. Dai}
\author{ W. Yu}
\affiliation{Department of Physics, Renmin University of China, Beijing 100872, China}

\author{B. P. Xie}\email{bpxie@fudan.edu.cn}
\author{D. L. Feng}\email{dlfeng@fudan.edu.cn}
\affiliation{State Key Laboratory of Surface Physics, Department of
Physics,  and Advanced Materials Laboratory, Fudan University,
Shanghai 200433, People's Republic of China}

\date{\today}

\begin{abstract}

The correlations between the superconductivity in iron pnictides and their electronic structures are elusive and controversial so far. Here through angle-resolved photoemission spectroscopy measurements, we show that such correlations are rather distinct in AFe$_{1-x}$Co$_x$As (A=Li, Na), but only after one realizes that they are orbital selective. We found that the superconductivity is enhanced by the Fermi surface nesting, but only when it is between $d_{xz}$/$d_{yz}$ Fermi surfaces, while for the $d_{xy}$ orbital, even nearly perfect Fermi surface nesting could not induce superconductivity. Moreover, the superconductivity is completely suppressed just when the $d_{xz}$/$d_{yz}$ hole pockets sink below Fermi energy and evolve into an electron pocket. Our results thus establish the orbital selective relation between the Fermiology and the superconductivity in iron-based superconductors, and substantiate the critical role of the $d_{xz}$/$d_{yz}$ orbitals. Furthermore, around the zone center, we found that the $d_{xz}$/$d_{yz}$-based bands are much less sensitive to impurity scattering than the $d_{xy}$-based band, which explains the robust superconductivity against heavy doping in iron-based superconductors.

\end{abstract}


\maketitle


The  orbital degree of freedom is responsible for many emergent properties in  correlated materials. For example in transition-metal oxides, due to the multiplicity of orbitals, the system
often exibits complex orbital and charge
orderings \cite{TMDreview}. In manganese oxides, $t_{2g}$ orbitals  are
strongly localized, while $e_g$ orbitals are itinerant. Such
different characters of different orbitals and the Hund's
coupling among them  lead to the colossal magnetoresistance
effect \cite{LMO}. In the iron-based high temperature superconductors (Fe-HTS's), the orbital degree of freedom also plays an important role. In the parent compounds,  a nematic electronic state has been discovered prior to the magnetic order, indicating the existence of orbital ordering and orbital fluctuations \cite{Matsuda, ZXreview, MYiBaCo}. In the superconducting state, the gap anisotropy, nodes, and multi-gap behaviors have been observed \cite{YZhangBaK, YZhangBaP}, which might be attributed to the different orbital compositions of  the Fermi surfaces (FSs).

Certain orbital dependencies of various properties in iron pnictides  are expected as for any multi-orbital system, however so far, the distinctive roles played  by various orbitals remain unclear.  Models including three, five or even more orbitals have been proposed to make quantitative calculations and comparisons \cite{ThreeBand, FiveBand}, however, they pratically conceal the dominating physics with complexity as well. In search of the critical correlations between superconductivity and electronic structures, at first,  the nesting between any hole FS at the zone center and any electron FS at the zone corner was suggested to be important for  the superconductivity, without distinguishing the orbital characters \cite{HDingBaK, HDingBaCo, FLNing}.  Later on,  it was pointed out  in BaFe$_{2-x}$Co$_x$As$_2$ \cite{Kaminski} and LiFeAs \cite{Borisenko} that FS nesting is unnecessary, while the presence of central hole pockets or Van Hove singularity are more important. Again,  the orbital degree of freedom was ignored in these studies, and the several hole pockets originated from different orbitals in BaFe$_{2-x}$Co$_x$As$_2$ were not resolved. On the other hand, the majority of theoretical studies  indicate that the inter-pocket nesting would significantly enhance the superconductivity \cite{KurokiT, Mazin}. Therefore, these contradicting correlations between the Fermi surface topology (or Fermiology for short)  and the superconductivity  create confusion, since they are all partially supported by different experiments and theories. It hampers a clear picture as to what determines the superconductivity in Fe-HTS's.

 In this Letter, we  examine the roles of  different orbitals in the superconductivity of Fe-HTS's with  angle resolved photoemission spectroscopy (ARPES) studies on LiFe$_{1-x}$Co$_{x}$As and NaFe$_{1-x}$Co$_{x}$As. With unpolar surface, and simple orbital characters of various FSs, these two systems are ideal for studying the orbital-selective role of the Fermiology in the superconductivity. We found the FS nesting effect on superconductivity is orbital selective;  even perfect nesting  for the $d_{xy}$ Fermi pockets in LiFe$_{0.83}$Co$_{0.17}$As could not induce superconductivity, while the perfect nesting between $d_{xz}$/$d_{yz}$ Fermi pockets corresponds to the maximal  superconducting transition temperature ($T_C$)  in  NaFe$_{1-x}$Co$_{x}$As.  In addition,  we found that superconductivity is completely suppressed when the $d_{xz}$/$d_{yz}$ hole pockets evolve into an electron pocket (a Lifshitz transition).  Furthermore, around the zone center, we observed that the $d_{xy}$-based band is much more susceptible to impurity scattering than the $d_{xz}$/$d_{yz}$-based bands. Our results give a distinct description of the roles of different orbitals, and establish a pivotal correlation between Fermiology and superconductivity, which could resolve the previous controversies and simplify  theories on the superconductivity.


\begin{figure}[t]
\includegraphics[width=8.5cm]{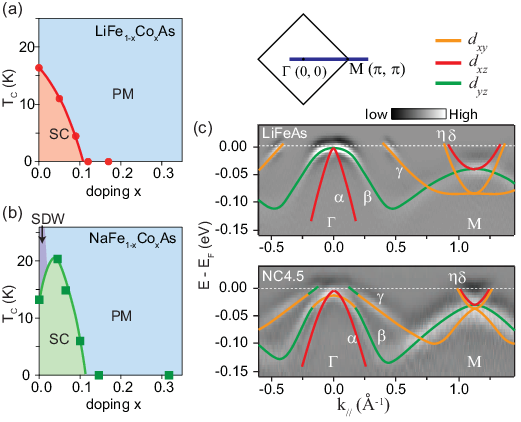}
\caption{(a) and (b), The phase diagrams of LiFe$_{1-x}$Co$_x$As and NaFe$_{1-x}$Co$_x$As, respectively. Transport measurements could be found in the supplementary information. (c) The second derivative photoemission spectra taken in LiFeAs and NC4.5 along the $\Gamma$-M direction. The orbital distribution and band structure are overlayed on the spectra. The same color code for different orbitals are used throughout this paper.}\label{transport}
\end{figure}

\begin{figure}[t]
\includegraphics[width=7.5cm]{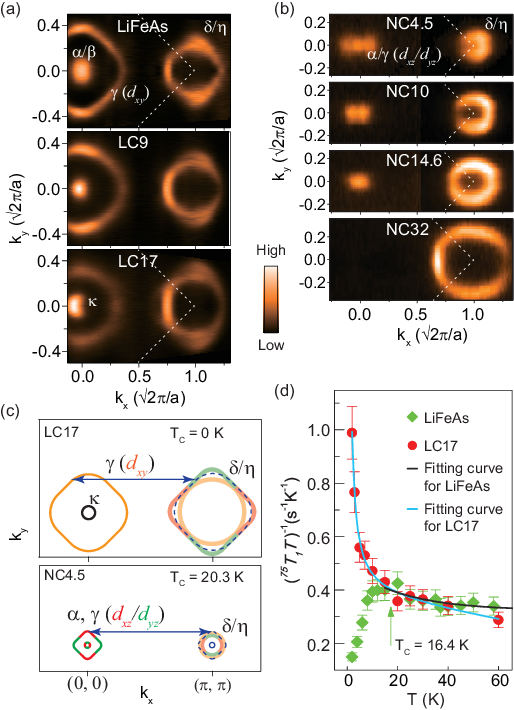}
\caption{(a) Doping evolution of the photoemission intensity map in
LiFe$_{1-x}$Co$_x$As taken with 21.2~eV photons in mixed polarization. (b) Doping evolution of the
photoemission intensity map in NaFe$_{1-x}$Co$_x$As taken with
100~eV photons in linear polarization. (c)
Illustration of the FS nesting condition in
LC17 and NC4.5. (d) The
spin-lattice relaxation rate 1/$^{75}$T$_1$T of LiFeAs and
LC17 under a field of 8~T and 11.5~T
respectively, applied along the crystal ab plane. The solid line is a
fit by a Curie-Weiss term 1/T$_1$T =A+B/(T-$\Theta$) with
$\Theta$=-20$\pm$5 K for LiFeAs and $\Theta$=-0.7$\pm$5 K for LC17. See ref. \cite{WQYu} for the NMR experimental details. }\label{mapping}
\end{figure}

High quality AFe$_{1-x}$Co$_{x}$As (A=Li, Na) single crystals of various dopings were synthesized with self-flux method (See supplementary material), covering a large portion of the phase diagrams [Figs.~\ref{transport}(a) and \ref{transport}(b)]. For LiFe$_{1-x}$Co$_{x}$As, superconductivity  was observed in x=0, 0.03, and 0.09 samples (named as LiFeAs, LC3, and LC9 hereafter by their dopant percentages) with $T_C$ of 16.4, 11, and 4.4~K, respectively [Fig.~\ref{transport}(a)]. For NaFe$_{1-x}$Co$_{x}$As,  the $T_C$'s are 13, 20.3, 14.8, and 6~K for x=0, 0.045, 0.065, and 0.1 respectively (named as NaFeAs, NC4.5, NC6.5, and NC10  hereafter) [Fig.~\ref{transport}(b)]. ARPES measurements were performed at Fudan University with a 7 eV Laser or 21.2 eV light from a helium discharging lamp, and also at various beamlines, including the beamline 5-4 of Stanford Synchrotron Radiation Lightsource (SSRL), the beamline 9A of Hiroshima Synchrotron Radiation Center (HiSOR) and the SIS beamline of Swiss Light Source (SLS). All the data were taken with Scienta R4000 electron analyzers. The
overall energy resolution was 5$\thicksim$10 meV at Fudan, SSRL and HiSOR, or 15$\thicksim$20
meV at SLS depending on the photon energy, and the angular
resolution was 0.3 degree. The samples were cleaved $\mathit{in~situ}$, and
measured in ultrahigh vacuum with pressure better than
3$\times$10$^{-11}$ torr.

The orbital characters and band structures of LiFeAs and NaFeAs have been well studied by previous ARPES  measurements \cite{Borisenko, MYiNaFeAs, YZhangNaFeAs}.  There are three hole bands near $\Gamma$ and two electron bands near M.  We compare the spectra taken along $\Gamma$-M direction for LiFeAs and NC4.5 as shown in Fig.~\ref{transport}(c). The band structures are very similar except for the $d_{xy}$ band. In LiFeAs, the $\gamma$ band with the $d_{xy}$ orbital crosses  the Fermi energy ($E_F$)  around the zone center forming a large hole pocket. However, in NC4.5, the $d_{xy}$ band disperses well below $E_F$ and hybridizes with the $d_{yz}$ band. Therefore, the $d_{xy}$ orbital does not contribute to the FS. Intriguingly,  $T_C$ is almost the same for LiFeAs and NC4.5, no matter whether the $d_{xy}$ orbital is present on the zone center FS  or not.

\begin{figure*}[t]
\includegraphics[width=16cm]{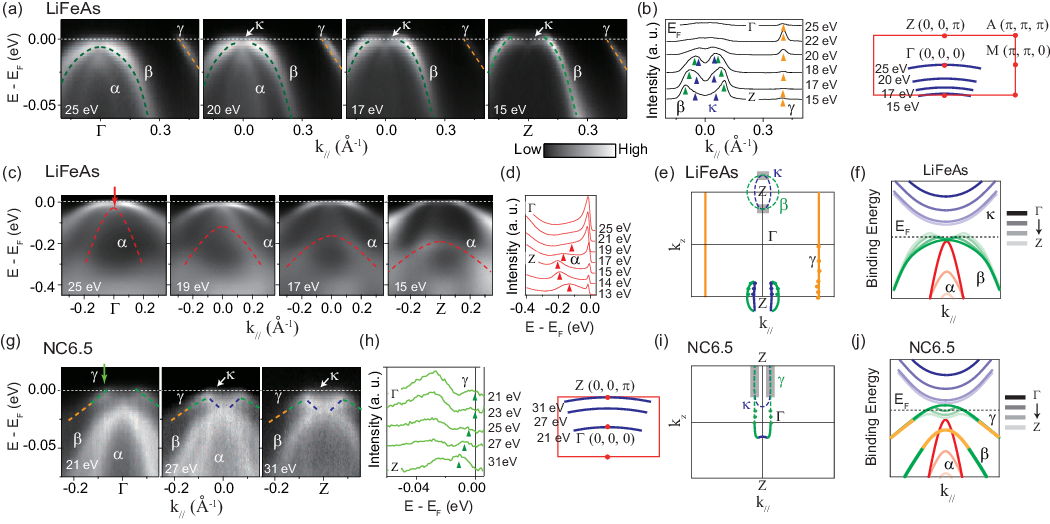}
\caption{Panels (a)~-~(f) are the $k_z$ dependent data for LiFeAs. (a) and (b), Photon energy dependence of the
photoemission spectra and the momentum distribution curves (MDCs) at $E_F$  taken along (0, 0)~-~($\pi$,
$\pi$) direction.  (c) and (d), Photon energy dependence of the photoemission spectra with a larger energy scale and the energy distribution curves (EDCs) at $k_{//}$~=~0 $\AA^{-1}$ taken along (0, 0)~-~($\pi$, $\pi$) direction. (e) Illustration of  the FS cross-section in
Z~-~$\Gamma$~-~M~-~A plane. The upper half of panel (e) shows the hybridization between $\beta$ and $\kappa$ pockets. Gap opens in the gray marked sections. (f) is a Cartoon showing how the bands evolute from $\Gamma$ (represented by darker color) to Z (represented by lighter color). (g)~-~(j) are the same as (a), and (d)~-~(f), respectively, but for NC6.5. The EDCs in (h) are taken at the Fermi crossings as indicated by the green arrow in panel (g).}\label{topology}
\end{figure*}

\begin{figure*}[t]
\includegraphics[width=16cm]{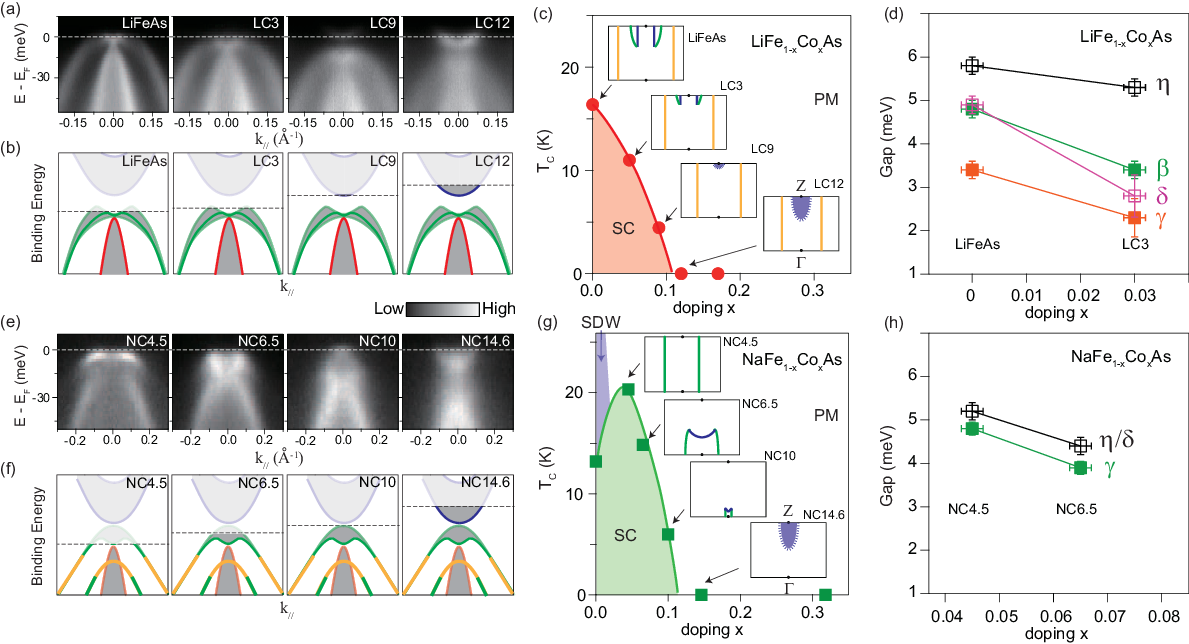}
\caption{(a) Doping dependence of the electronic structure around the zone center in  LiFe$_{1-x}$Co$_x$As, taken with a 7eV laser. (b) is illustrations of the bands in (a). (c) The phase diagram and corresponding FS topology for LiFe$_{1-x}$Co$_x$As near zone center.  (d) The doping dependence of superconducting gap in LiFe$_{1-x}$Co$_x$As. The gap sizes are determined by symmetrized EDC's taken on different FSs (See supplementary information).  (e)~-~(h) are the same as (a)~-~(d), but for NaFe$_{1-x}$Co$_x$As. The spectra in panel (e) are taken with 31~eV photons. In panels (c) and (g), the solid lines represent hole FSs, while dashed ones with blue area inside represent electron FSs.}\label{band}
\end{figure*}

Replacing Fe with Co introduces electrons  into the system. As shown in Figs.~\ref{mapping}(a) and \ref{mapping}(b), the hole pockets shrink with Co doping while the electron pockets are enlarged.  According to the spin-fluctuation-mediated pairing scenario \cite{KurokiT, Mazin} , similar sizes of hole and electron pockets will give better nesting and thus enhance the scattering from zone center to zone corner, which will benefit the superconducting pairing. For LiFe$_{1-x}$Co$_{x}$As, the nesting condition for the $d_{xy}$ hole pocket is improved with Co doping. However, the $T_C$ is suppressed. We could even achieve a nearly perfect nesting between the $d_{xy}$ hole pocket and the electron pockets in LC17, but it is not superconducting at all [the upper panel of Fig.~\ref{mapping}(c)].  On the contrary, in NC4.5 the $d_{xz}$/$d_{yz}$ hole pockets are well nested with the electron pockets, and its $T_C$ is the highest one in this series [the lower panel of Fig.~\ref{mapping}(c)].  With further electron doping, the nesting worsens and T$_C$ decreases [Fig.~\ref{mapping}(b)].  The comparison between  LiFe$_{1-x}$Co$_{x}$As and NaFe$_{1-x}$Co$_{x}$As indicates that the FS nesting does enhance superconductivity, but only when it is between $d_{xz}$/$d_{yz}$ FSs. This is consistent with the larger superconducting gaps on the $d_{xz}$/$d_{yz}$ FSs here shown  in Figs.~\ref{band}(d) and \ref{band}(h), and for Ba$_{0.6}$K$_{0.4}$Fe$_2$As$_2$ as well \cite{HDingBaK, BorisenkoBK}.  Moreover, we compare the strength of spin fluctuations in LiFeAs and LC17 measured by the nuclear magnetic resonance (NMR).  The fitting parameter $\Theta$ (Curie-Weiss temperature) of LC17 is much larger than that of LiFeAs, indicating stronger low-energy spin fluctuations in LC17 [Fig.~\ref{mapping}(d)]. However, it is important to note  that $T_C$ is not enhanced by  such an  enhancement of the low-energy spin fluctuations due to the nesting of  the $d_{xy}$ FSs.

\begin{figure}[t]
\includegraphics[width=8cm]{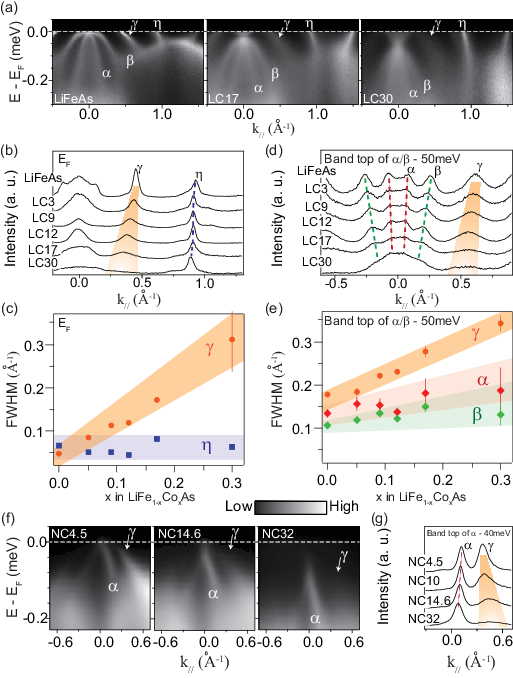}
\caption{(a) Doping dependence of the photoemission intensities of LiFe$_{1-x}$Co$_x$As. (b) The doping dependence of the MDCs at $E_F$.  (c) is the corresponding FWHMs of $\gamma$ and $\eta$ in (b).  (d)  MDCs at 50~meV below the band top of $\alpha$/$\beta$ as a function of doping, since $\alpha$ and $\beta$ do not cross $E_F$. (e) is the corresponding FWHMs of $\alpha$, $\beta$ and $\gamma$ in (d).  (f) Doping dependence of the photoemission intensities in NaFe$_{1-x}$Co$_x$As. (g) Doping dependence of the MDCs at 40~meV below the band top of $\alpha$ in NaFe$_{1-x}$Co$_x$As.}\label{impurity}
\end{figure}

To further study the Fermiology in 3D momentum space, Fig.~\ref{topology}(a) shows the $k_z$ evolutions of the bands around the zone center in LiFeAs, obtained by changing the incident photon energy (top right inset of Fig.~\ref{topology}).  From $\Gamma$ to Z, the band top of $\beta$ gradually shifts toward $E_F$, and finally crosses $E_F$ around  Z. Meanwhile, an electron-like band named $\kappa$ emerges at the photon energy of $\thicksim$20~eV, and hybridizes with $\beta$, exhibiting a ``M''-like feature near $E_F$. The Fermi crossings of $\beta$, $\kappa$, and $\gamma$ could be traced clearly from the  MDCs near $E_F$ [Fig.~\ref{topology}(b)]. The observation of $\kappa$ electron pocket is notable. According to band calculations, there is a fast dispersing electron band along the $k_z$ direction whose band bottom is far above $E_F$ at $\Gamma$, but shifts downward quickly when approaching Z \cite{LDALiFeAs}. It has been proposed that when the bottom of this band touches $\alpha$/$\beta$, the  top of $\alpha$ will be pushed downwards away from $E_F$ \cite{Kuroki}. Consistently, in Figs.~\ref{topology}(c) and \ref{topology}(d), the  top of $\alpha$ shifts downwards quickly from $\Gamma$ to Z.  Such a strong $k_z$ dispersion of $\alpha$ and the appearance of $\kappa$ produce a distinctive 3D Fermiology for the $d_{xz}$/$d_{yz}$ hole pocket. As shown in Fig.~\ref{topology}(e), $\beta$ forms an ellipsoidal hole FS along the $k_z$ direction, while the $\kappa$ electron pocket appears around Z and hybridizes with the $\beta$ hole pocket. As a result, the energy gaps open on the crossings of the two FS sheets, forming two banana-like shaped FS cross-sections. The $k_z$ evolution of the band structure of LiFeAs is illustrated in Fig.~\ref{topology}(f). Note that, the finite $k_z$ resolution of ARPES will smear out the bands with strong $k_z$ dispersions.  As shown in Fig.~\ref{topology}(c),  the band dispersion of $\alpha$ at $\Gamma$ still contributes finite spectra weight at 19~eV and 17~eV.  This is perhaps why such a strong $k_z$ dispersion of this band was missed in previous ARPES experiments.

For  NC6.5, as shown in Fig.~\ref{topology}(g), the $\kappa$ electron pocket emerges at 27~eV and opens a hybridization gap  with the $\gamma$ band [Fig.~\ref{topology}(h)], which resembles that in LiFeAs. However,  the 3D FS topology is very different in NC6.5.  As shown in Figs.~\ref{topology}(i) and \ref{topology}(j) , the $\gamma$ band contributes a cylindrical hole FS from $\Gamma$ to Z. When the $\kappa$ band appears around Z,  $\kappa$ and $\gamma$ cross and opens a hybridization gap, and consequently the FS evolves into a  drum-like hole pocket surrounding $\Gamma$.

Now we  examine how these  $d_{xz}$/$d_{yz}$-based 3D FSs evolve with doping in LiFe$_{1-x}$Co$_x$As and NaFe$_{1-x}$Co$_x$As.  Figure~\ref{band}(a) demonstrates the doping evolution of the low lying electronic structures in LiFe$_{1-x}$Co$_x$As measured by a 7~eV Laser. With Co doping, all the bands shift downwards, and the $\kappa$ electron pocket could be clearly observed in LC9 and LC12. Note that, the $\beta$ band with a strong $k_z$ dispersion is broadened or even split at $E_F$ due to the finite $k_z$ resolution or final state effect. This suggests that the $k_z$ broadening effect should also be carefully considered even in laser-based ARPES experiments.  We thus compare the spectra with the experimentally-determined  band structure integrated from $\Gamma$ to Z  in Fig.~\ref{band}(b). Figure~\ref{band}(c) illustrates the corresponding 3D FS topology for different doping levels.  With Co doping, the banana-like hole pockets sink away from $E_F$ and the $\kappa$ electron pocket emerges around Z in LC12 sample.  Similar phenomena could be also observed in NaFe$_{1-x}$Co$_{x}$As  [Figs.~\ref{band}(e)~-~\ref{band}(g)]. The   $d_{xz}$/$d_{yz}$ FS is a cylinder in NC4.5 and shrinks with Co doping. In NC14.6, $\kappa$ electron pocket emerges. The FS evolution of LiFe$_{1-x}$Co$_x$As and NaFe$_{1-x}$Co$_x$As clearly shows that the $d_{xz}$ /$d_{yz}$  FS undergoes a Lifshitz  transition.  Although the Fermiology is dramatically different for  LiFe$_{1-x}$Co$_x$As and NaFe$_{1-x}$Co$_x$As, we found in both cases that the superconductivity disappears just quickly after the disappearance  of the $d_{xz}$/$d_{yz}$ hole FSs [Figs.~\ref{band}(c) and \ref{band}(g)]. Our results thus suggest the importance of the presence of central $d_{xz}$/$d_{yz}$ hole FSs  for the superconductivity in Fe-HTS's. Such an orbital selective correlation between Lifshitz transition and superconductivity is resolved for the first time. In addition, we note that LC9 is still superconducting below 5K, while the small $\kappa$ FS has appeared at  $E_F$ [Fig.~\ref{band}(c)].  It is likely due to some residual superconducting pairing amplitude contributed by the rest of the FSs, after the $d_{xz}$/$d_{yz}$ hole FSs sink below $E_F$. When the $\kappa$ FS grows bigger at higher doping,  it does not help the superconductivity. In the inter-pocket pairing scenario,  it is known that the pairing strength would be weak if the Fermi velocities at the zone corner and zone center hold the same sign.  That is, superconductivity would not be enhanced when both are electron FSs around the zone center and corner, as is the case for our data. Alternatively, the suppression of superconductivity could be viewed through the decreasing superconducting gap [Figs.~\ref{band}(d), \ref{band}(h), and supplementary information], which is roughly proportional to $T_C$.

The orbital selective correlation between superconductivity and the electronic structure is also manifested  in the impurity scattering effects. As shown in Fig.~\ref{impurity}(a), with Co doping, the $d_{xy}$-based $\gamma$ band becomes significantly weaker and broader. Figs.~\ref{impurity}(b)~-~\ref{impurity}(e) plot the MDCs at $E_F$ and 50~meV below the band top of the $\alpha$/$\beta$ bands, together with the full-width-half-maximum (FWHM) of various bands. The FWHM of $\gamma$ increases remarkably with Co doping comparing to all the other bands. On the other hand, because the band top of $\alpha$/$\beta$  shifts away from $E_F$, the increasing binding energy causes the slight increase of FWHMs with doping for the $\alpha$ and $\beta$ bands in Fig.~\ref{impurity}(e). Therefore, the FWHMs of $\eta$, $\alpha$ and $\beta$ are essentially insensitive to the Co dopants. Similar phenomena is also observed in NaFe$_{1-x}$Co$_x$As [Figs.~\ref{impurity}(f) and \ref{impurity}(g)].  Such  orbital-selective scattering effects of the Co dopants need further theoretical understandings. However, it might explain the robust superconductivity against heavy doping in Fe-HTS's, since most bands with $d_{xz}$ and $d_{yz}$ orbitals are basically unaffected by the scattering of dopants. Furthermore, a recent STM study of NaFe$_{1-x}$Co$_x$As shows that the low energy electronic state is somehow insensitive to the Co-dopant \cite{HHWen}.  Our data provide an explanation: the tunnelling matrix element might be weak for the in-plane $d_{xy}$ orbital that is most sensitive to the impurity scattering.

The correlations that we found here for iron pnictides between the superconducitivity and the hole FSs around the zone center do not  apply to  K$_x$Fe$_{2-y}$Se$_2$  and  the single FeSe layer on  SrTiO$_3$ substrate \cite{YZhangKFeSe,STOFeSe,DLFeSe}. They both have high $T_C$'s but no  hole pockets. Scattering between the electron pockets around the zone corner was suggested to be important for the superconductivity in these iron selenides \cite{dwave}.  Interestingly, the FS of NC32 is almost the same  as  that of K$_x$Fe$_{2-y}$Se$_2$ [Fig.~\ref{mapping}(b)], but NC32 is not superconducting. Our results indicate that the superconducting mechanism of these iron selenides are remarkably different from that of iron pnictides. Other factors may come into play, for example,  the lattice constants of these iron selenides are much larger than those of iron pnictides \cite{DLFeSe}.


To summarize, our results substantiate the pivotal role of the $d_{xz}$/$d_{yz}$ orbitals in the superconductivity of AFe$_{1-x}$Co$_x$As (A=Li, Na), and establish the orbital selective correlation between the superconductivity and the Fermiology. Although the Fermiology is dramatically different in these two series, we demonstrate that the superconducting $T_C$ is maximized only by the perfect nesting  between $d_{xz}$/$d_{yz}$-originated FSs, while the superconductivity  diminishes quickly after the central $d_{xz}$/$d_{yz}$ hole FSs disappear with electron doping.
Moreover, we found that the $d_{xz}$/$d_{yz}$  orbitals are rather insensitive to impurity scattering.
Our results thus could be used to simplify theories for the superconductivity in iron pnictides \cite{JP1, JP2}, and help to design new materials to enhance $T_C$ based on the Fermiology. It also clarifies many previous contradicting statements on the correlations.

We gratefully acknowledge the experimental support by Dr. D. H. Lu, Dr. R. G. Moore at SSRL, Dr. M. Arita at HiSOR and Dr. M. Shi at SLS. This work is supported in part by the National Science Foundation of China, and National Basic Research Program of China (973 Program) under the grant Nos. 2012CB921400, 2011CBA00112. SSRL is operated by the US DOE, Office of BES, Divisions of Chemical Sciences and Material Sciences.

\end{document}